\documentstyle[aps,manuscript,epsfig]{revtex}
\newcommand{\beq}{\begin{equation}}
\newcommand{\eeq}{\end{equation}}
\newcommand{\ba}{\begin{array}}
\newcommand{\ea}{\end{array}}

\begin{document}
\title{Ultra high energy cosmic rays from extragalactic astrophysical sources:
energy losses and spectra}
\author{V. Berezinsky $^{1,3}$, A.Z. Gazizov $^{2}$ and S.I. Grigorieva$^{3}$}
\address{$^{1}$INFN, Laboratori Nazionali del Gran Sasso,
I-67010 Assergi (AQ), Italy,\\
$^{2}$B. I. Stepanov Institute of Physics of the National
Academy of Sciences of Belarus,\\
F. Skariny Ave.\ 68, 220062 Minsk, Belarus\\
$^{3}$Institute for Nuclear Research of Russian Academy of Sciences,\\ 
60th Anniversary of October Revolution prospect 7A, 117312 Moscow, Russia}
\maketitle

\begin{abstract}
The energy losses and spectra of Ultra High Energy Cosmic Rays (UHECR) are
calculated for protons as primary particles. The attention is given to the
energy losses due to electron-positron production in collisions with the
microwave 2.73 K photons. The energy spectra are calculated for several
models, which differ by production spectra and by source distribution,
namely: {\it (i)} Uniform distribution of the sources with steep generation
spectra with indices 2.4 - 2.7, with cosmological evolution and without it.
In this case it is possible to fit the shape of the observed spectrum up to $%
8\cdot 10^{19}$~eV, but the required CR emissivity is too high and the GZK
cutoff is present.  
{\it (ii)} Uniform distribution of the sources with flat
generation spectrum $dE/E^2$ which is relevant to GRBs. The calculated
spectrum is in disagreement with the observed one. The agreement at $%
E\lesssim 8\cdot 10^{19}$~eV can be reached using a complex generation
spectrum, but the required CR emissivity is three orders of magnitude higher
than that of GRBs, and the predicted spectrum has the GZK cutoff. {\it (iii)}
The case of local enhancement within region of size 10 - 30 Mpc with
overdensity given by factor 3 - 30. The overdensity larger than 10 is
needed to eliminate the GZK cutoff.
\end{abstract}

\section{Introduction}

The energy losses of UHE protons in extragalactic space are caused by
interaction with microwave radiation. The contribution of IR and optical
radiation is small (for a detailed review of energy losses and the resulting
spectrum see \cite{book}). The main contribution to energy losses is given
by expansion of the universe, electron-positron pair production and pion
production. The latter process results in steepening of the proton spectrum
referred to as the Greisen-Zatsepin-Kuzmin (GZK) cutoff \cite{GZK}.  The GZK
cutoff is not seen in the observational data (for a recent review see \cite
{NaWa}). The most conservative approach to explanation of observations is
astrophysical one: the protons are accelerated in extragalactic
astrophysical sources (normal galaxies, compact objects in normal galaxies,
e.g. GRB engines, AGN etc) and propagate towards us. This approach comprises
three aspects: acceleration to UHE, total energy release in a source and
propagation in extragalactic space. This most conservative approach is
considered as (almost) excluded, with certain caveats, however. The models
in which the GZK cutoff is absent or ameliorated include nearby {\it %
one-source model} (see \cite{WW,BeDoGr} and most recent work \cite{Bier});
the {\it Local Supercluster model}, in which the density of UHECR sources is
locally enhanced (\cite{book,BeGrLS}, for a recent work see \cite{BlBlOl});
and finally widely discussed {\it GRB model} which, according to
calculations \cite{Wa}, gives a reasonable agreement with observations.

In this paper we shall analyse the two latter models.

\section{Energy losses}

We present here the accurate calculations for pair production,
$p+\gamma_{CMBR}\to p+e^++e^-$, and for pion production $p+\gamma_{CMBR}\to N+%
{\rm pions}$, where $\gamma_{CMBR}$ is a microwave photon.

The energy losses of UHE proton per unit time due to its interaction
with low energy photons is given by
\beq 
-\frac{1}{E}\frac{dE}{dt}=\frac{c}{2\Gamma^2}\int_{\epsilon_{th}}^{\infty}
d\epsilon_r\sigma(\epsilon_r)f(\epsilon_r)\epsilon_r\int_{\epsilon_r/2\Gamma}%
^{\infty} d\epsilon \frac{n(\epsilon)}{\epsilon^2},
\label{enloss}
\eeq
where $\Gamma$ is the Lorentz factor of the proton, $\epsilon_r$ is
the energy of background photon in the system where the proton is at rest,
$\epsilon_{th}$ is the threshold of the considered reaction in the rest
system of the proton, $\sigma(\epsilon_r)$ is the cross-section, 
$f(\epsilon_r)$ is the mean fraction of energy lost by the proton in
one $p\gamma$ collision in the laboratory system, $\epsilon$ is the
energy of the background photon in the lab system, and $n(\epsilon)$
is the density of background photons. 

For the CMBR with temperature $T$ Eq.(\ref{enloss}) is simplified  
\beq
-\frac{1}{E}\frac{dE}{dt}=\frac{cT}{2\pi^2\Gamma^2}\int_{\epsilon_{th}}%
^{\infty}d\epsilon_r\sigma(\epsilon_r)f(\epsilon_r)\epsilon_r
\left\{-\ln\left[1-\exp\left(-\frac{\epsilon_r}{2\Gamma
T}\right)\right]
\right\}.
\label{CMBRloss}
\eeq
From Eqs.(\ref{enloss}) and (\ref{CMBRloss}) one can see that the
mean fraction of energy lost by the proton in lab system in one
collision, $f(\epsilon_r)=\langle 1-x \rangle=(E_p-E^{\prime}_p)/E_p$, 
is the basic quantity needed for 
calculations of energy losses. The threshold values of these quantities
are well known: 
\beq
f_{\rm pair}\approx \frac{2m_e}{m_p},~~~ ~~~~~
f_{\rm pion}\approx \frac{\epsilon_r}{m_p}\frac{1+\mu^2/2\epsilon_r m_p}
{1+2\epsilon_r/m_p},
\label{frac}
\eeq
where $f_{\rm pair}$ and 
$f_{\rm pion}$ are the threshold fractions for 
$p+\gamma \rightarrow p+e^++e^-$ and $p+\gamma \rightarrow N+\pi$, 
respectively, and $\mu$ is the pion mass. 

For the accurate calculations of energy losses the fraction $f$
properly averaged over differential cross-section is needed. 

Pair production loss has been previously discussed in many papers. All
authors directly or indirectly have followed the standard approach of Ref.~
\cite{Blumenthal} where the first Born approximation of the Bethe-Heitler
cross-section with proton mass $m_p\rightarrow\infty$ was used. In contrast
to Ref.~\cite{Blumenthal}, we are using the first Born approximation
approach 
of Ref.~\cite{BergLinder}, which takes into account the finite proton mass. We
also use the exact non-relativistic differential cross-section
from \cite{LL}. 
This allowed us to calculate the average fraction of energy lost by
the proton in lab system  by performing fourfold integration over 
invariant mass of electron-positron pair $M_X$, over an angle between
incident and scattered proton, and polar and azimuthal angles of an
electron in the c.m system  of the pair (see Appendix A for further details).


Calculating photopion energy loss we followed the method of papers
\cite{BeGa,GNPCS}. Total cross-sections were taken according to Ref.~\cite
{totcrs}. At low c.m.\ energy $E_c$ we considered the binary reactions 
$p+\gamma \rightarrow \pi+ N$ (including the resonance $p+\gamma \rightarrow
\Delta$),~~ $p+\gamma \rightarrow \pi^- +\Delta^{++}$, and $p+\gamma
\rightarrow \rho^0+p$. Differential cross-sections of binary processes at
small energies were taken from \cite{pioncrs}. At $E_c > 4.3$~GeV we assumed
the scaling behaviour of differential cross-sections. These were taken from
Ref.~\cite{scal}. In the intermediate energy range we used an interpolation
approach which allows us to describe the residual part of the total 
cross-section.
The corresponding differential cross-sections coincide with low-energy
binary description and high-energy scaling distribution and have a smooth
transition between these two regimes in the intermediate region. 
\begin{figure}[htb]
\epsfxsize=12truecm
\centerline{\epsfbox{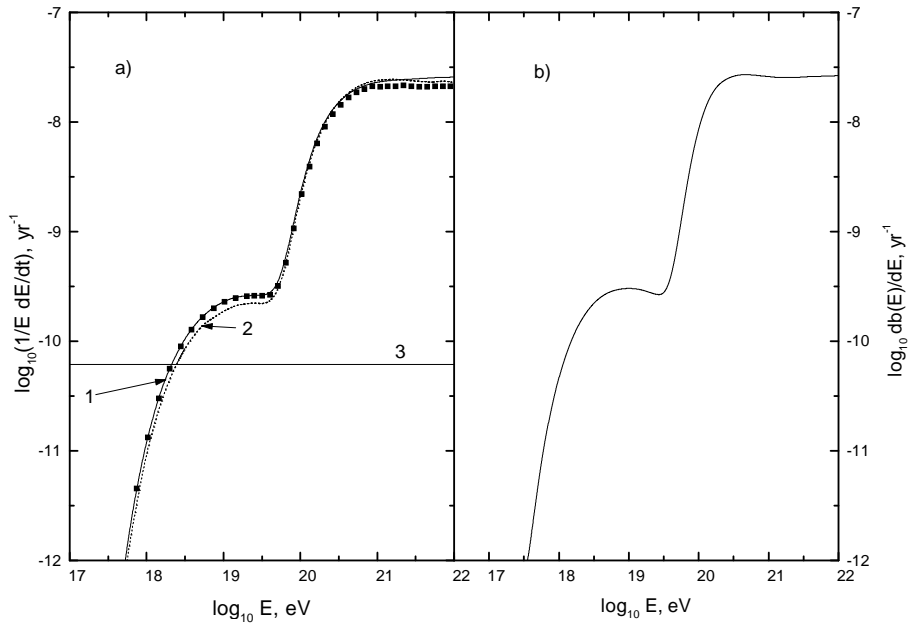}}
\vspace{2mm}\caption{a){\em UHE proton energy losses $E^{-1}dE/dt$ (present work:
curve 1; Berezinsky and Grigorieva (1988) \protect\cite{BeGrbump}: curve 2;
Stanev et al 2000 \protect\cite{stanev} : black squares). The line 3 gives
energy losses due to redshift ($H_0=65$~ km/sMpc). b) The derivative
$db_0(E)/dE$, where $b_0(E)=dE/dt$ at present epoch $z=0$.}}
\label{loss}
\end{figure}
\noindent
The results
of our calculations are presented in Fig.\ref{loss} in terms of relative
energy losses per unit time $E^{-1}dE/dt$ as function of energy (curve 1).
Also plotted is the derivative $db_0(E)/dE$, where $b_0(E)=dE/dt$
(Fig.~\ref{loss}b). This
quantity is needed for calculation of differential energy spectrum (see
section III). In Fig.~\ref{loss} we plot for comparison the energy losses as
calculated by Berezinsky and Grigorieva 1988 \cite{BeGrbump} (dashed curve
2). The difference in energy losses due to pion production is very small,
not exceeding 5\% in the energy region relevant for comparison with
experimental data($E\leq 10^{21}~eV$). The difference with energy losses due
to pair production is larger and reaches maximal value 15\%. The results of
calculations by Stanev et al \cite{stanev} are shown by black squares. These
authors have performed the detailed calculations for both aforementioned
processes, though their approach is somewhat different from ours, especially
for photopion process. Our energy losses are practically indistinguishable
from \cite{stanev} for pair production and low energy pion production, and
differ by 15-20\% for pion production at highest energies (see Fig.~\ref
{loss}). Stanev et al claimed that energy losses due to pair production is
underestimated by Berezinsky and Grigorieva \cite{BeGrbump} by 30-40\%.
Comparison of {\it data files} of Stanev et al and Berezinsky and Grigorieva
(see also Fig.\ref{loss}) shows that this difference is significantly less.
Most probably, Stanev et al scanned inaccurately the data from the journal
version of the paper \cite{BeGrbump}.

\section{Uniform distribution of UHECR sources and GZK cutoff}

\begin{figure}[htb]
\epsfxsize=10truecm
\centerline{\epsffile{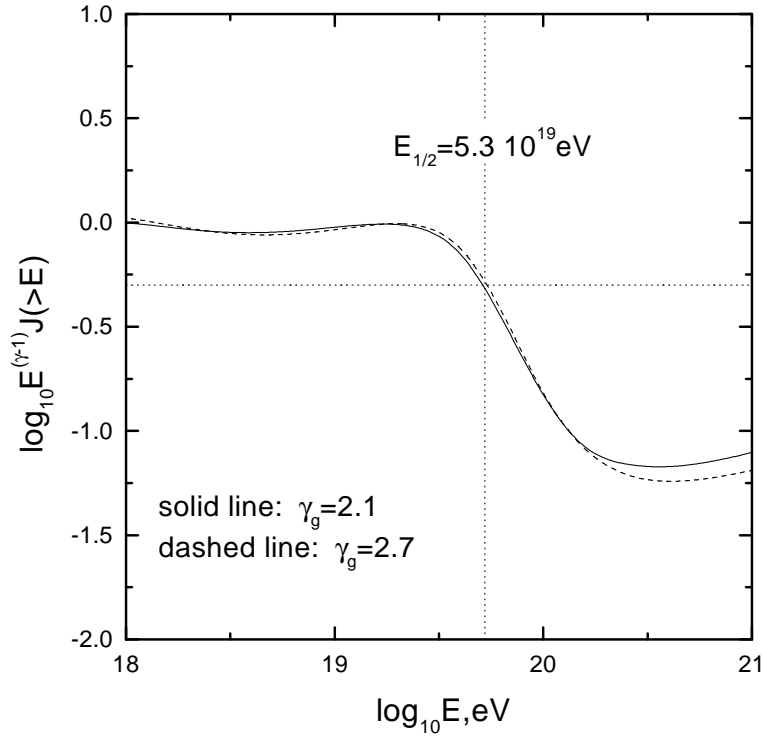}}
\vspace{1mm}
\caption{{\em Integral UHECR spectra with indices of generation spectra $%
\protect\gamma_g=2.1$ and $\protect\gamma_g=2.7$ (solid and dashed line,
respectively). The vertical dotted line shows the energy $E_{1/2}$, where
the calculated flux becomes two times lower than power-law extrapolation.}}
\label{Ehalf}
\end{figure}

The GZK cutoff is a model--dependent feature of the spectrum, e.g. the GZK
cutoff for a single source depends on the distance to the source. A common
convention is that the GZK cutoff is defined for diffuse flux from the
sources uniformly distributed over the universe. In this case one can give
two definitions of the GZK cutoff. In the first one the cutoff is determined
as the energy, $E_{GZK} \approx 3\times 10^{19}~eV$, where the steep
increase in the energy losses starts (see Fig.~\ref{loss}). The GZK cutoff
starts at this energy. The corresponding pathlength of a proton is $%
R_{GZK}\approx (E^{-1}dE/dt)^{-1}\approx 1.3\cdot 10^3$~Mpc. The advantage
of this definition of the cutoff energy is independence on spectrum index,
but this energy is too low to judge about presence or absence of the cutoff
in the measured spectrum. More practical definition is $E_{1/2}$, where the
flux with cutoff becomes lower by factor 2 than power-law extrapolation.
This definition is convenient to use for the integral spectrum, which is
better approximated by power-law function, than the differential one. In Fig.%
\ref{Ehalf} the function $E^{(\gamma-1)}J(>E)$ , where $J(>E)$ is calculated
integral diffuse spectrum, is plotted as function of energy. Note, that $%
\gamma>\gamma_g$ is an effective index of power-law approximation of the
spectrum modified by energy losses. For wide range of generation indices $%
2.1 \leq \gamma_g \leq 2.7$ the cutoff energy is the same, $E_{1/2} \approx
5.3\cdot 10^{19}$~eV. The corresponding proton pathlength is $R_{1/2}
\approx 800$~Mpc.

Using energy losses given in Section 2, we calculated the diffuse spectra
for the model when sources are distributed uniformly in the universe. We
followed the method of calculation suggested in Ref.\cite{BeGrbump}.
We use two assumptions for uniform distribution of the sources: {\it (i)}
with evolution of the sources described by factor $(1+z)^m$ in comoving
frame \cite{book}, and {\it (ii)} without evolution. The power-law
generation spectrum with index $\gamma_g$ was assumed. We made different
assumptions about maximum energy in the generation spectrum, namely $E_{{\rm %
max}}=3\cdot 10^{20}$~eV~~, $E_{{\rm max}}=1\cdot 10^{21}$~eV, and $E_{{\rm %
max}}=\infty$. Varying parameters $\gamma_g$ and $m$ we fit the AGASA and
Akeno data \cite{Teshima}.

The fit of UHECR data with help of evolving sources was made in the past
(e.g. see Ref.\cite{BeGr83} and \cite{book}). The widely used fit for the
AGASA data with $\gamma_g=2.3$ and with assumed mixed composition of
galactic and extragalactic UHECR was  found by Yoshida and 
Teshima \cite{Teshima}. Recently Scully and Stecker \cite{Stecker}
made calculations similar to that above for UHECR produced by GRBs.

We calculate spectra using the formalism of Ref.\cite{BeGrbump}:
\begin{equation}
J_p(E)=(\gamma_g-2)\frac{c}{4\pi}\frac{{\cal L}_0}{H_0}E^{-\gamma_g}
\int_0^{z_{max}}dz_g(1+z_g)^{m-5/2}\lambda^{-\gamma_g}(E,z_g) 
\frac{dE_g(z_g)}{dE},  
\label{flux}
\end{equation}
where $z_g$ is a redshift at generation and $E_g(z_g)$ is energy of a proton
at generation, if at present ($z=0$) its energy is $E$: $E_g(z_g)=
\lambda(E,z_g)E$ and $\lambda(E,z_g)$ is calculated numerically using energy
losses $dE/dt$ accounted for their time evolution; ${\cal L}_0=n_0 L_p$ is CR
emissivity at $z=0$ ($n_0$ and $L_p$ are space density of the sources and
their CR luminosity, respectively). As the general case we assume
cosmological evolution of the sources given by 
${\cal L}(z)={\cal L}_0(1+z)^m$, where the absence of evolution
corresponds to $m=0$. All energies in Eq.(\ref{flux}) are given in GeV
and luminosities in GeV/s. Dilation of energy
interval is given by \cite{BeGrbump} (see also Appendix B):
\begin{equation}
\frac{dE_g(z_g)}{dE}=(1+z_g)\exp\left[\int_0^{z_g}\frac{dz}{H_0}(1+z)^{1/2}
\left( \frac{db_0(E^{\prime})}{dE^{\prime}}\right)_{E^{\prime}=(1+z)E_g(z_g)}
\right],  
\label{dilation}
\end{equation}
where $b_0(E)=dE/dt$ is energy loss due to interaction with CMBR photons at 
$z=0$ (adiabatic energy loss due to redshift must not be included!). 
Derivative $db_0(E)/dE$ at $z=0$ is given in Fig.(\ref{loss}b).

For particles with energies $E\gtrsim 1\times 10^{17}$~eV the maximum
redshift  for evolution of CR sources $z_{\rm max}$ is not important
if it is larger than 4. 
Integration over large z gives small contribution  
when the generation energy $E_g(E,z)$ reaches the value $E_{\rm eq}(z_m)$,
at which energy losses due to pair production and redshift are equal.
Then the maximum redshift $z_m(E)$ of the epochs contributing to the 
flux of protons with energy E is determined by equation  
$(1+z_m)E=E_{\rm eq}(z_m)$.
For energies $E< 1\times 10^{17}$~eV the maximum redshift of the 
source evolution $z_{\rm max}$  might be important. In these cases we fix it
as $z_{\rm max}=4$.
\begin{figure}[htb]
\epsfxsize=10truecm
\centerline{\epsffile{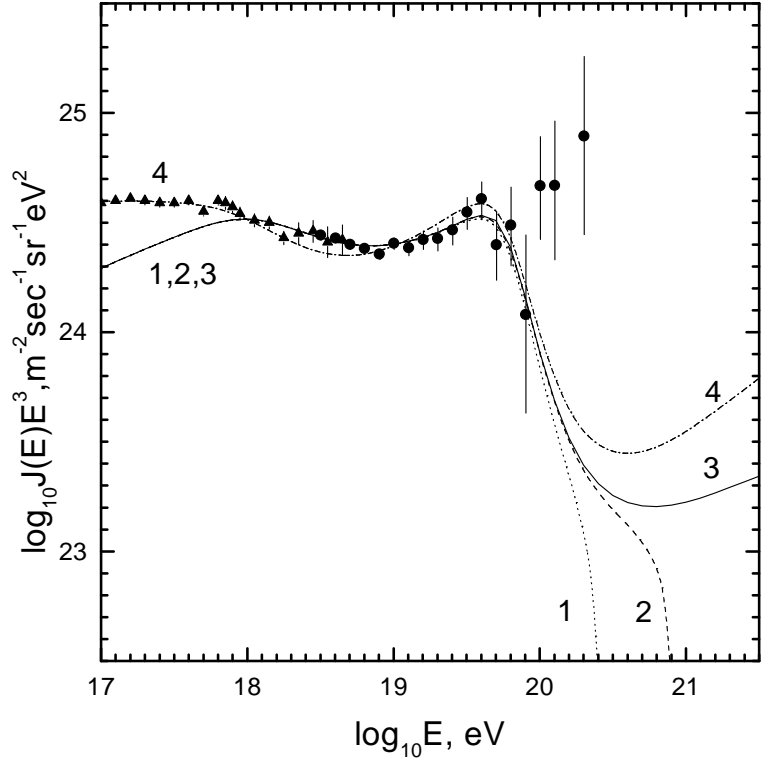}}
\vspace{2mm}
\caption{{\em UHECR spectrum as observed in Akeno (triangles) and AGASA
(filled circles) experiments. The curves show the predicted differential spectra
for the uniform distribution of sources with or without evolution. The case
without evolution ($m=0$, $\protect\gamma_g=2.7$) is given by curves
(1),(2),(3) for maximum generation energy $E_{max}= 3\cdot 10^{20}~ {\rm eV}%
,~ 1\cdot 10^{21}~ {\rm eV}$ and $\infty$, respectively. The dashed
curve 4
describes the evolutionary model with $m=4$,~~
$\protect\gamma_g=2.45$ and $E_{\rm max}=\infty$.}}
\label{spectra}
\end{figure}
We can fit the Akeno-AGASA data in both cases, with and without evolution.
The spectra without evolution, $m=0$ can fit the data starting from
relatively high energy $E\geq 1\cdot 10^{18}$~eV. The fit needs $\gamma_g=2.7
$. The curves 1, 2 and 3 in Fig.\ref{spectra} show the spectra with
different $E_{max}$ equal to $3\cdot 10^{20}~ {\rm eV},~ 1\cdot 10^{21}~{\rm %
eV}$ and $\infty$, respectively. The fit without evolution (curves 1,~2,~3)
needs ${\cal L}_0=4.7\cdot 10^{51}$~erg/Mpc$^3$yr, while the
fit for evolutionary case (curve 4) needs ${\cal L}_0=1.3\cdot 10^{49}$%
~erg/Mpc$^3$yr. The difference
between these two emissivities is caused mainly by flatter generation
spectrum in the evolutionary case.

The required emissivities can be compared with most powerful local
emissivity given
by Seyfert galaxies ${\cal L}_{Sy}=n_{Sy}L_{Sy}$. Using the space density of
Seyfert galaxies $n_{Sy} \sim 10^{-77}$~cm$^{-3}$ and the luminosity $L_{Sy}
\sim 10^{44}$~erg/s one obtains ${\cal L}_{Sy} \sim 1\cdot 10^{48}$~erg/Mpc$%
^3$yr, which is almost 4 orders of magnitude less than CR emissivity needed
in no-evolutionary case and one order of magnitude less than one in the
evolutionary case.

As Fig.\ref{spectra} shows the models with uniform distribution of the
sources are excluded by absence of GZK cutoff in the observations. They give
a good fit to the lower energy data. This fit needs large $\gamma_g$ and thus
too large energy output of the sources, $nL$. It is possible to
overcome this difficulty using an  assumption that production spectrum
is flat at low energies and has a steepening at some high energy
$E_c$. Assuming, for example,  that spectrum is $\propto E^{-2}$ at
low energy, and $\propto E^{-2.7}$ at $E\geq E_c=1\times 10^9$~GeV, one
obtains the required CR emissivity ( see Section V)
${\cal L}=3.7\times 10^{46}$~erg/Mpc$^3$yr, i.e. less than observed
total emissivity due to the Seyfert galaxies. A plausible assumption is
that the population of UHECR sources is comprised by galaxies with
moderate activity of AGN, which at higher luminosities are linked to
Seyfert galaxies and BL Lacertae. There were recently found the
observational indications that the latter galaxies could be the sources
of observed UHECR \cite{Tkach}. If such sources had large local
overdensity, the GZK cutoff would be less noticeable. We shall
study this possibility in the next Section.

\section{Local overdensity of UHECR sources}

Local overdensity of UHECR sources makes the GZK cutoff less sharp or
eliminates it \cite{book}. Clustering of galaxies is a gravitational
property, which is determined by mass and not by internal activity of an
object. The galaxies of the same masses with active galactic nuclei
or without them, with burst of star formation or in quiet phase,
are clustering in the same way. Therefore the optical catalogues give a
reasonable indication to expected clustering of UHECR sources.

\begin{figure}[htb]
\epsfxsize=10truecm
\centerline{\epsffile{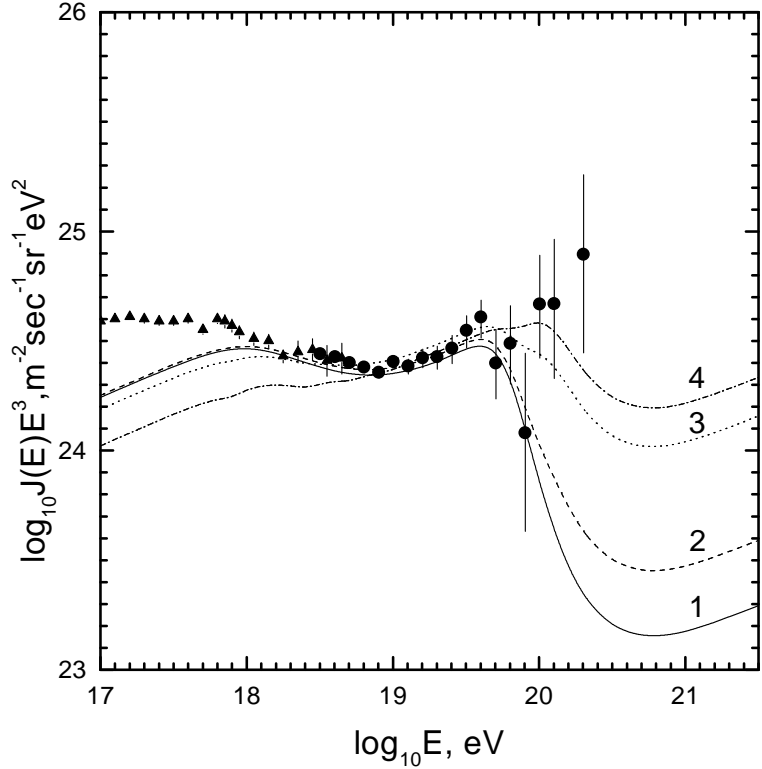}}
\vspace{2mm}
\caption{{\em The effect of source overdensity on UHECR spectra for
$R_{overd}=30~Mpc$ and different overdensity ratios $n/n_0=1,~ 2,~ 10,~ 30$
(curves 1,~ 2,~ 3 and 4, respectively).}}
\label{LS}
\end{figure}
\noindent
The nearby
structure that can affect the GZK cutoff is Local Supercluster (LS) of
galaxies, which has a form of ellipsoid with semi-axes 20 and 30 Mpc. The LS
overdensity of galaxies is estimated by factor $\sim 2$ ( see \cite{Peebles}
and references therein). Such overdensity does not solve the problem of GZK
cutoff \cite{BeGrLS,BlBlOl}. We shall calculate here UHECR spectra for
different local overdensities $n/n_0$, where $n_0$ is mean extragalactic
density of UHECR sources. We use the various sizes of overdensity region $%
R_{overd}$, equal to 10, 20 and 30 Mpc. The results of our calculations are
presented in Fig.\ref{LS} for $\gamma_g=2.7$, ~ $m=0$ and for four values
of overdensity $n/n_0$ equal to 1,~ 2,~ 10 and 30, assuming the size of
overdensity region 30~Mpc (the results for $R_{overd}=20~$Mpc are not much
different). From Fig.\ref{LS} one can see that overdensity $n/n_0 \gtrsim 10$
is needed to reconcile well the calculations with observational data.

\section{UHECR from GRB}

In GRBs the protons can be accelerated to Ultra High Energies 
\cite{Vietri,Wax95}. The strong indication that UHECR can be produced by GRBs,
the authors of Ref.\cite{Vietri,Wax95,Wa} see in the equal emissivity ${\cal %
E}$ in GRBs and UHECRs. Scully and Stecker \cite{Stecker} argue that in fact
the energy output in cosmic rays is higher than in GRBs. We shall analyse
here the problem of energy output combined with the spectrum shape.

\begin{figure}[htb]
\vspace{.5cm}
\epsfxsize=10truecm
\centerline{\epsffile{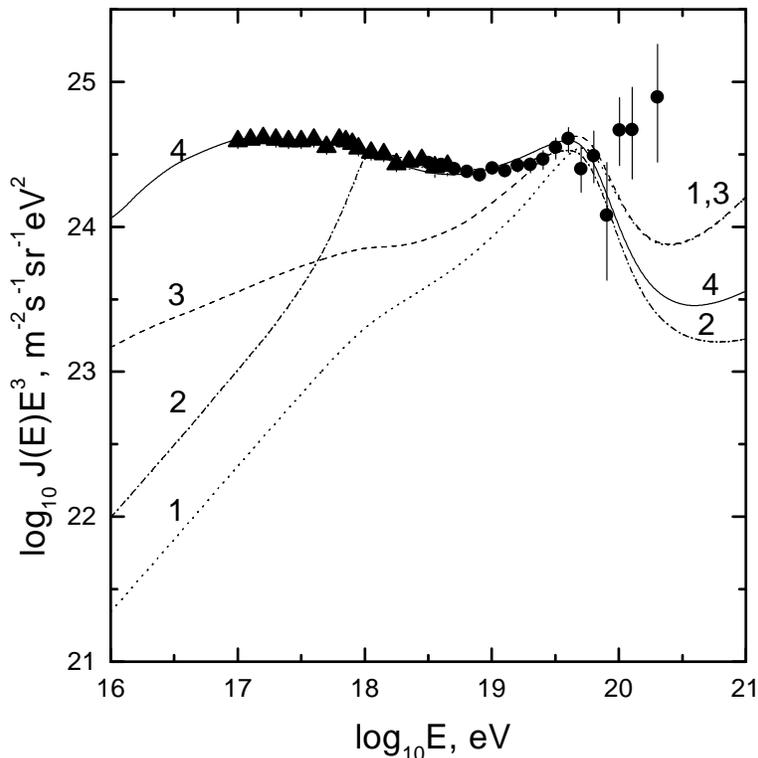}}
\caption{{\em UHECR spectra from GRBs for $E^{-2}$ generation spectrum 
(curve 1 for non-evolutionary $m=0$ case and curve 3 for evolutionary $m=4$ 
case) and for $E^{-2}$ spectrum with steepening (curve 2 for $m=0$,~
$E_c=1\cdot 10^9$~GeV,  $\gamma_g=2.7$, and curve 4 for $m=4$,~
$E_c=1\cdot 10^8$~GeV, $\gamma_g=2.45$)}}
\label{GRB}
\end{figure}

For energetically most favourable CR generation spectrum $dE/E^2$, advocated
in \cite{Vietri,Wax95}, the diffuse spectrum of UHECR can be found as
\begin{equation}
J_p(E)=\frac{c}{4\pi}~\frac{1}{\ln\frac{E_{max}}{E_{min}}}~ \frac{{\cal L}_0%
}{H_0}E^{-2} \int_0^{z_{max}}dz_g(1+z_g)^{m-5/2}\lambda^{-2}(E,z_g)\frac{dE_g%
}{dE}. 
\label{GRBs}
\end{equation}
The calculated spectra for non-evolutionary case $m=0$ and for evolution of
GRB sources with $m=4$ are displayed in Fig.\ref{GRB} by curves 1 and 3,
respectively. The required CR emissivity is 
${\cal L}_0=2.0\cdot 10^{45}$~erg/Mpc$^3$yr for both cases. It is two orders
of magnitude larger than that observed in GRBs \cite{Schmidt} 
${\cal E}_{GRB}=1\cdot 10^{43}$~erg/Mpc$^3$yr.

Apart from the problem of too large energy output, these models do not fit
the observed spectrum shape and predict the standard GZK cutoff. To obtain
the agreement with spectrum shape one can use an artificial $E^{-2}$
spectrum with steepening at energy $E_c$.

At energy $E>E_c$ the generation spectrum of a source is
\begin{equation}
Q_g(E_g,z)=\frac{L_p(z)E_c^{\gamma_g-2}}{\ln\frac{E_c}{E_{min}}+\frac{1} {%
\gamma_g-2}}E_g^{-\gamma_g},  \label{gen}
\end{equation}
while at $E<E_c$ this spectrum is assumed to have $1/E^2$ shape. It is easy
to verify that this spectrum is correctly normalized to the luminosity 
$L_p$. The diffuse spectrum can be readily calculated at $E\geq E_c$ as
\begin{equation}
J_p(E)=\frac{cH_0^{-1}}{4\pi}~{\cal L}_0~\frac{E_c^{\gamma_g-2}E^{-\gamma_g}} 
{\ln\frac{E_c}{E_{min}}+\frac{1}{\gamma_g-2}} \int_0^{z_{max}}
dz_g(1+z_g)^{m-5/2} \lambda^{-\gamma_g}(E,z_g)\frac{dE_g}{dE}.  
\label{GRBspectr}
\end{equation}
The fluxes given by Eq.(\ref{GRBspectr}) are
displayed in Fig.\ref{GRB} by curves 2 and 4 for cases $m=0$ and $m=4$,
respectively. These spectra agree well with the Akeno-AGASA observations at
energies lower than the GZK cutoff, but require higher emissivity, 
${\cal L}_0=3.7\cdot 10^{46}$~erg/Mpc$^3$yr ($m=0$ case, curve 2) and 
${\cal L}_0=3.1\cdot 10^{46}$~erg/Mpc$^3$yr ($m=4$ case, curve 4).

We conclude thus that UHECR from GRBs exhibit the standard GZK cutoff and
require CR emissivity 2 -- 3 orders of magnitude higher than that observed
in GRBs. Our conclusions agree with that of Ref.\cite{Stecker}.

\section{Conclusions}

We have performed accurate calculations of energy losses of UHE protons due to
electron-positron pair production and pion production in collisions with
microwave photons. The diffuse spectra of UHE protons have been calculated
for uniform distribution of the sources in the universe for different
maximum energies of the generation spectrum. The generation spectrum with
index $\gamma_g=2.7$ provides a good fit for energy range $1\cdot 10^{18}
- 8\cdot 10^{19}$~eV in case of absence of evolution. For the case of
evolution the good fit is given by $m=4$ and $\gamma_g=2.45$ in the energy
range $1\cdot 10^{17} -8\cdot 10^{19}$~eV. 

Local overdensity of UHECR sources, e.g. in Local Supercluster, can
reconcile the weak GZK cutoff with UHECR data only if overdensity is larger
than 10. The existing astronomical data favour much smaller overdensity, of
order of 2.

The excellent fit of observational data in energy range 
$1\cdot 10^{17} - 8\cdot 10^{19}$~ eV by astrophysical spectrum is
rather impressive. It requires too high local CR emissivity 
${\cal L}_0 \sim 1\cdot 10^{49}$~erg/Mpc$^3$yr (in case of an
evolutionary model), but this emissivity can be drastically reduced
down to $3\cdot 10^{46}$~erg/Mpc$^3$yr by assumption of $E^{-2}$
spectrum with steepening at $E_c=1\cdot 10^8$~GeV. Normal galaxies
with weak AGN ({\em quasi-seyferts}) meet this energy requirement.
The local overdensity of these galaxies could ameliorate the GZK
problem, but required overdensity, $n/n_0 \geq 10$, is larger than
that observed $n/n_0 \approx 2$.

UHECR from GRBs have a standard GZK cutoff. The predicted $E^{-2}$
generation spectrum gives bad fit to the observed spectrum at energy lower
than GZK cutoff. In case the generation spectrum is modified to give a
reasonable fit, the required CR emissivity exceeds that observed in GRBs by
three orders of magnitude.

\section*{Acknowledgements}

We are grateful to Todor Stanev who provided us with data file of energy
losses calculated in Ref.\cite{stanev}. We thank M.Teshima for sending
us the AGASA data for zenith angles $\theta \leq 45^{\circ}$ which were
used in these calculations (the data for $\theta \leq 60^{\circ}$ are 
still preliminary).  

The work was partially supported by
INTAS through grant INTAS 99-1065 and  by Russian Foundation for Basic
Research grant 00-15-96632.

\appendix\section{Calculations of energy losses}

Pair production energy loss of ultrahigh-energy protons in
low-energy photon gas, e.g. \emph{CMBR},
\beq
 p + \gamma_{CMBR} \rightarrow p + e^+ + e^-
\label{eep}
\eeq
has been previously discussed in many papers. The differential
cross-section for this process in the first Born approximation was
originally calculated in 1934 by Bethe and Heitler \cite{BetHeit}
and Racah \cite{Racah}. In 1948 Feenberg and Primakoff
\cite{FeenPrim} obtained the pair production energy loss rate
using the extreme relativistic approximation for the differential
cross-section. And in 1970 the accurate calculation was performed
by Blumenthal \cite{Blumenthal}. Later some analytical
approximations to differential cross-sections were applied to this
problem in Ref.~\cite{Chodor}.

All authors neglected the recoil energy of proton putting
$m_p\rightarrow\infty$, the effect being suppressed by a factor of
$m_e/m_p \approx 5 \times 10^{-4}$.

In spite of the fact that all calculations actually used the same
Blumenthal approach, there are noticeable discrepancies in the results
of different authors; they were clearly demonstrated in Fig.~1b of
the Ref.~\cite{stanev}.

To clarify the situation we recalculated the pair production
energy loss of high-energy proton in the low-energy photon gas.
In contrast to Ref.~\cite{Blumenthal} we use the first Born
approximation approach of Ref.~\cite{BergLinder} taking into
account the finite proton mass. The exact non-relativistic
threshold formula with corrections to different Coulomb
interactions of electron and positron with the proton (see e.g.\
Ref.~\cite{LL}) was used. No series expansions of
$\sigma(E_\gamma)$ were involved in our calculations.

Our strategy was to calculate the average energy transfer
$x=E^\prime_p/E_p$, where $E_p$ and $E^\prime_p$ are the incident and
final proton energies respectively, in the laboratory system by
performing the direct fourfold integration of the exact matrix
element over the phase space. It should be noted, that direct
numerical integration, especially at high energies, is 
difficult in this case because of forward-backward spikes in the
electron-positron angular distributions. To overcome this problem,
we performed two integrations over polar and azimuth angles in the
$e^+e^-$ subsystem analytically. This was facilitated by using of
the \emph{MATHEMATICA 4} code. The residual two integrations over
energy and scattering angle in the initial $p\gamma$ subsystem
were carried out numerically. We calculate simultaneously
the total cross-section for pair production. The
accuracy of our calculations was thus controlled  by comparison of 
calculated total cross-section with the well-known Bethe-Heitler 
cross-section.

\begin{figure}[htb]
\epsfxsize=11truecm
\centerline{\epsffile{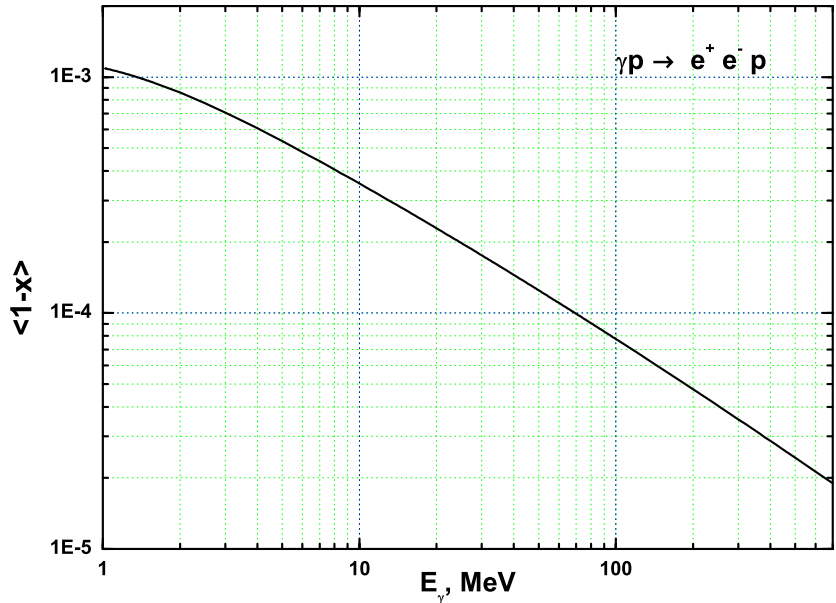}}
\caption{\em The average fraction of the incident proton energy
$E_p$ carried away by $e^+e^-$ pair as a function of the photon
energy $E_\gamma$ in the proton rest system.}
\label{1-x}
\end{figure}

The average fraction of proton energy lost in one collision with a photon is
plotted in Fig.~\ref{1-x} as a function of the photon energy in
the proton rest system.

\begin{figure}[htb]%
\epsfxsize=11truecm
\centerline{\epsffile{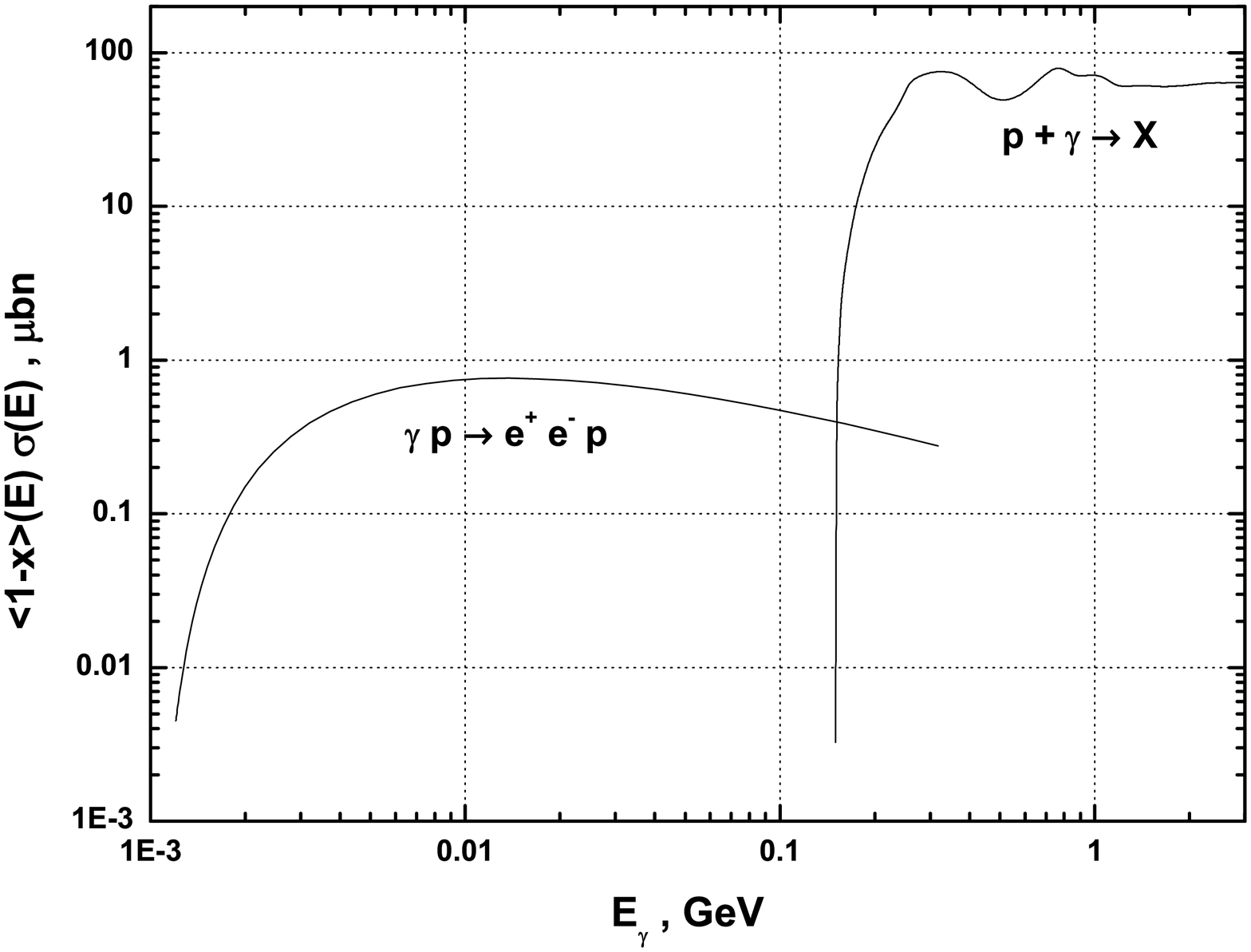}}
\caption{\em The product of fraction of energy lost, $\langle 1-x \rangle$, 
and the cross-section for pair production,
$\gamma p \rightarrow e^+e^-p$, or photopion production 
$p+\gamma \rightarrow X$ as function of photon energy in the proton
rest system $E_{\gamma}$.} 
\label{comp}
\end{figure}

The product of this fraction and the total cross-section for pair production 
is shown in Fig.~\ref{comp}. This function should be integrated over
the photon spectrum  to obtain the average energy loss due to pair production
in the photon gas with this spectrum.


The comparison of our calculations with  Ref.~\cite{stanev} shows the
negligible difference (see Fig.~\ref{loss}).


\section{Connection between energy intervals at epochs of
production and observation }

If we consider the protons with energy $E$ in the interval $dE$
at the epoch with redshift $z=0$, what will be the corresponding interval
of generation $dE_g$ at epoch $z$, when energy of a proton was $E_g(z)$? 
The connection between these two intervals is given by Eq.(36) of 
Ref.\cite{BeGrbump}. Here we shall confirm this formula using a
different, more simple derivation. Note, that intermediate formulae 
(40) and (41) used for derivation of final Eq.(36) in Ref.\cite{BeGrbump}
have a misprint: the correct power of $(1+z)$ term there is 3, not 2. 

Regarding the energy losses of a proton on CMBR, $dE/dt$ and $(1/E)dE/dt$, at
arbitrary epoch with redshift $z$, we shall use the following notation
\beq
b(E,z)= \left (\frac{dE}{dt}\right)_{{\rm CMBR}},~ ~ ~ 
\beta(E,z)=\left( \frac{1}{E}\frac{dE}{dt}\right )_{{\rm CMBR}}.
\label{notation}
\eeq
As it is readily seen from Eq.(\ref{CMBRloss}), the energy losses
at arbitrary epoch $z$ can be found as 
\beq 
\beta(E,z)= (1+z)^3\beta_0\left ((1+z)E\right ),
\label{zloss}
\eeq
where notation $b_0(E)$ and
$\beta_0(E)$  are used  here and henceforth for energy losses at $z=0$. 
The energy loss due to redshift at the epoch $z$ is given by
\beq
\left (\frac{1}{E}\frac{dE}{dt}\right )_{{\rm r-sh}}=H_0(1+z)^{3/2}.
\label{r-sh}
\eeq
The energy of a particle at epoch $z$,
\beq
E_g(z)= E+ \int_t^{t_0}dt\left [ \left (\frac{dE}{dt}\right )_{{\rm r-sh}}
+\left (\frac{dE}{dt}\right )_{{\rm CMBR}}\right],
\label{E-g,t}
\eeq
can be easily rearranged as 
\beq
E_g(z)=E+\int_0^z\frac{dz'}{1+z'}E_g(z')+\frac{1}{H_0}\int_0^z
\frac{dz'}{(1+z')^{1/2}}b_0\left ((1+z')E_g(z')\right ),
\label{E-g,z}
\eeq
with help of Eq.(\ref{zloss}) and expression $dt=dz/H_0(1+z)^{5/2}$ for time
interval.

Differentiating Eq.(\ref{E-g,z}) over $E$,  one finds for energy
interval dilation $y(z) \equiv dE_g(z)/dE$:
\beq
y(z)=1+\int_0^z\frac{dz'}{1+z'}y(z')+\frac{1}{H_0}\int_0^z
dz'(1+z')^{1/2}y(z')\left (\frac{db_0(E')}{dE'}\right )_{E'=(1+z')E_g(z')}.
\label{eq-z}
\eeq
Correspoinding differential equation is
\beq
\frac{1}{y(z)}\frac{dy(z)}{dz}=\frac{1}{1+z}+\frac{1}{H_0}(1+z)^{1/2}
\left (\frac{db_0(E')}{dE'}\right )_{E'=(1+z)E_g(z)}.
\label{eq-z,diff}
\eeq
The solution of Eq.(\ref{eq-z,diff}) is 
\beq
y(z)\equiv \frac{dE_g(z)}{dE}=(1+z)\exp\left[ \frac{1}{H_0}\int_0^z 
dz'(1+z')^{1/2}\left( \frac{db_0(E')}{dE'}\right )_{E'=(1+z')E_g(z')}\right],
\label{dilation1}
\eeq
where $E_g(z)$ is an energy at epoch z. Eq.(\ref{dilation1}) coincides
with Eq.(36) from Ref.\cite{BeGrbump}.

\end{document}